\documentclass[%
 aip,
 amsmath,amssymb,
 reprint,%
]{revtex4-1}

\usepackage{graphicx}
\usepackage{dcolumn}
\usepackage{bm}

\usepackage[utf8]{inputenc}
\usepackage[T1]{fontenc}
\usepackage{mathptmx}
\usepackage{etoolbox}
\usepackage{xcolor}

\definecolor{rev0}{rgb}{0.0, 0.0, 0.0}      

\makeatletter
\def\@email#1#2{%
 \endgroup
 \patchcmd{\titleblock@produce}
  {\frontmatter@RRAPformat}
  {\frontmatter@RRAPformat{\produce@RRAP{*#1\href{mailto:#2}{#2}}}\frontmatter@RRAPformat}
  {}{}
}%
\makeatother
\begin{document}

\preprint{AIP/123-QED}

\title[Cold plasma singularities]{Singularities of the cold plasma theory: 
Modeling challenges for ICRF operation in low-density edge plasma}
\author{W. Tierens}
 \email{tierenswv@ornl.gov}
\author{C. Klepper}%
\affiliation{ 
Oak Ridge National Laboratory, 1 Bethel Valley Road,
Oak Ridge, TN 37830, USA
}%

\author{R. Diab}
 
\affiliation{%
MIT Plasma Science and Fusion Center, Cambridge, Massachusetts 02139, USA
}%

\author{G. Urbanczyk}
\affiliation{%
Institut Jean Lamour UMR 7198 CNRS-Universit\'e de Lorraine, 2 all\'ee Andr\'e Guinier, F-54011 Nancy, France
}%

\date{\today}

\begin{abstract}
Sustained ICRF operation in a fusion power plant may require low edge densities to mitigate plasma-wall interactions, a regime which was recently achieved in WEST with very little impurity sputtering. Cold plasma theory, however, predicts singular radiofrequency electric fields in this regime, both at the lower hybrid resonance and along the resonance cones, raising the question of whether standard collisional cold plasma models suffice to describe low-density edge ICRF at all. Collisions in principle remove these singularities, replacing them with finite but sharply peaked fields. We derive these peak length scales analytically and confirm them with a 2D finite-element simulation using exponential mesh refinement, achieving $\mu$m resolution where needed. We conclude that edge collisions in cold plasma do not remove the need to resolve length scales ordinarily associated with hot-plasma and Bernstein-wave physics\footnote{This manuscript has been authored by UT-Battelle, LLC, under Contract No. DE-AC05-00OR22725 with the U.S. Department of Energy (DOE). This material is based upon work supported by the U.S. Department of Energy, Office of Science, Office of Advanced Scientific Computing Research and Office of Fusion Energy Sciences, Scientific Discovery through Advanced Computing (SciDAC)
program. The publisher acknowledges the US government license to provide public access under the DOE Public Access Plan (http://energy.gov/downloads/doe-publicaccess-plan).}.
\end{abstract}

\maketitle

Ion Cyclotron Resonance Frequency (ICRF) heating is one of the most promising heating systems for a future fusion power plant. Sustained operation of an ICRF heating system may require operating the ICRF antenna at low plasma densities, in order to avoid plasma-material interactions. 
Cold plasma theory predicts that low-density operation presents risks: in addition to poor coupling, very high radiofrequency electric fields are predicted to form in this regime, potentially causing breakdown or excessive sputtering. Yet this obstacle appears to be surmountable, as demonstrated on WEST \cite{Diab2026NF,diab2026direct}: almost no PMI impurities are produced in this regime.
This provides confidence in innovative antenna designs which would operate at low densities, such as a Traveling Wave Antenna \cite{hillairet2026design}.

Numerically modeling ICRF operation at these low densities is therefore of considerable importance. While at higher densities, Finite Element calculations have provided impressively accurate results, both for the coupled power \cite{suarez2020validation} and for the impurity sputtering \cite{kumar2025integrated}, in this low density scenario, the singularities of the cold plasma theory remain a fundamental problem, both numerically (``Can standard numerical Finite Element schemes handle these fields?'') and physically (``Is collisional cold plasma even the right electromagnetic description of the edge plasma?'').

Singular electric fields exist in the collisionless cold plasma theory at the Lower Hybrid Resonance \cite{campos2017constructive}, and along the resonance cones \cite{tierens2024slow,paulus2025icrf}. In both cases, the cold plasma theory nonetheless predicts a finite amount of power is coupled to the singular wave mode \cite{tierens2026power,maquet2021analytical}, even in the collisionless limit. 

The standard way to introduce collisions in the dielectric tensor is by modifying particle masses according to collision frequencies $\nu_e,\nu_i$ for the electrons and ions:
\begin{align}
    m_{e,eff}&= m_e(1+i \nu_e/\omega) \\
    m_{i,eff}&= m_i(1+i \nu_i/\omega) 
\end{align}
The cyclotron frequency $q B/m$ and plasma frequency $\sqrt{n q^2/m/\epsilon_0}$ are then calculated based on these complex effective masses, and the dielectric tensor is computed from those. 
The collision frequencies $\nu_e,\nu_i$ may be calculated from the Coulomb collision frequency \cite{hinton1983collisional} ($\nu \propto n T^{-3/2}$) for elastic electron-ion collisions, or from tabulated cross sections for electron-neutral and ion-neutral collisions.

\begin{figure}
    \centering
    \includegraphics[width=0.49\linewidth]{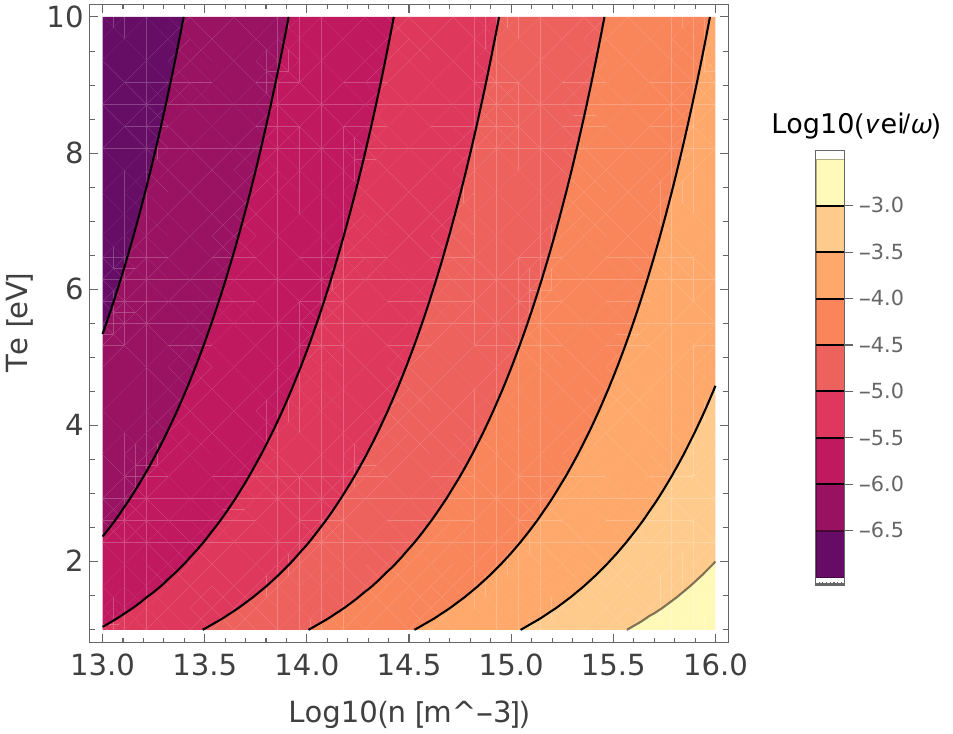}
    \includegraphics[width=0.49\linewidth]{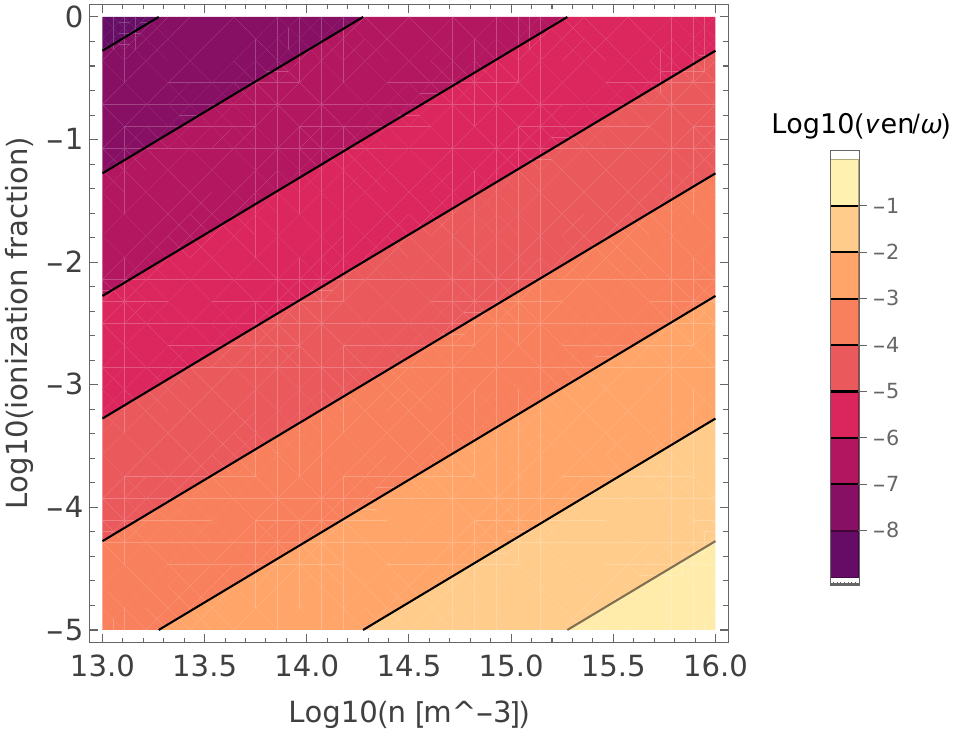}
    \caption{Left: $\log_{10}(\nu_{ei}/\omega)$ for Coulomb electron-ion collisions at relevant densities and temperatures, and $\omega=2\pi \cdot (6 \text{ MHz})$. Right: $\log_{10}(\nu_{en}/\omega)$ for electron-neutral collisions at relevant densities and ionization fractions ($n_{\text{plasma}}/n_{\text{neutral}}$), for a neutral species with a typical reaction rate of $10^{-13}$m$^3$/s.}
    \label{fig:collisions}
\end{figure}

From \cite{tierens2026helicon}, elastic electron-neutral collisions for neutral deuterium has reaction rates of order $10^{-13}$m$^3/$s. Achieving MHz collision frequencies \textcolor{rev0}{(at which $\nu/\omega\sim1\%$)} therefore requires an edge neutral density of $10^{19}$m$^{-3}$. At $n_e=10^{16}$m$^{-3}$, a density typical for the Lower Hybrid resonance, this would require the ionization fraction $n_e/n_n$ to be of order \textcolor{rev0}{$10^{-3}$}.
Other neutral impurities, e.g. Oxygen, do not appear to have substantially larger cross sections for electron-neutral elastic collisions \cite{song2026cross}. This is shown in figure \ref{fig:collisions}.
The reaction rate for ion-neutral charge exchange collisions, $p+H\rightarrow H+p$, is \textcolor{rev0}{an order of magnitude lower\cite{stangeby2000plasma}, $\sim 10^{-14}$m$^3/$s. If the neutral species is molecular H$_2$ or D$_2$ (rather than atomic H or D), the ion-neutral interaction is dominated by Langevin polarization capture\cite{pinto2008three}, with a reaction rate that is another order of magnitude lower $\sim 10^{-15}$m$^3/$s}.

The Coulomb collision frequency for elastic electron-ion collisions,
\begin{align}
    \nu_{ei}&=\frac{n_i Z^2 e^4 \ln \Lambda}{6\pi^2\epsilon_0^2 m_e^2 v_{th,e}^3}
\end{align}
is shown in figure \ref{fig:collisions}.
With ICRF-typical wave frequency $\omega=2\pi f$, $f$ on the order of tens of MHz, from figure \ref{fig:collisions} we see that even $\nu_e/\omega=1/100$ is hard to achieve, requiring very low temperatures $\sim 1$ eV or very low ionization fractions ($n_{\text{plasma}}/n_{\text{neutral}}\sim 10^{-3}$).

\begin{figure}
    \centering
    \includegraphics[width=0.75\linewidth]{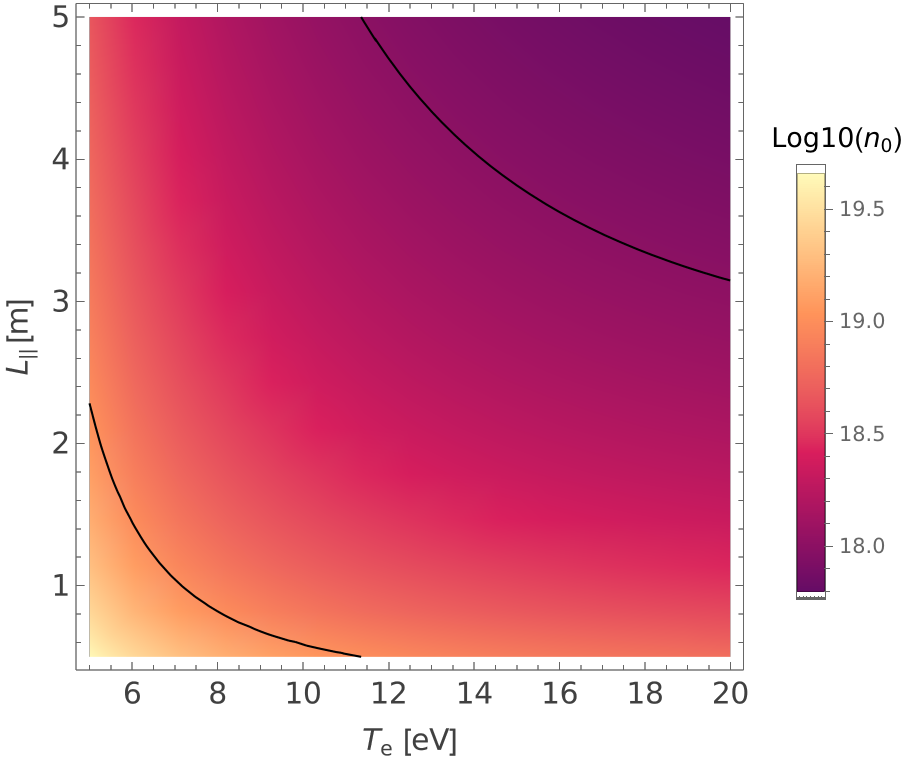}
    \caption{Estimated neutral density in a private SOL with characteristic connection length $L_{\parallel}$.}
    \label{fig:ionfrac}
\end{figure}

\textcolor{rev0}{How plausible are such low ionization fractions? Were the plasma in thermodynamic equilibrium, the Saha equilibrium\cite{fridman2008plasma} suggests near-complete ionization at $T_e>10$eV. In the private SOL, the plasma is not in thermodynamic equilibrium: it is optically thin, three-body recombination is weak, and plasma is rapidly lost along the magnetic field
to nearby material structures. To zeroth order, we must instead estimate the neutral density from an equilibrium between ionization sources and transport losses:
\begin{align}
    n_e n_0 K_{i}(T_e)=N_{\text{end}}\frac{n_i c_s}{L_{\parallel}} \label{estn0}
\end{align}
where $n_0$ is the neutral density, $n_i\approx n_e$, $c_s$ is the ion speed of sound, $N_{\text{end}}=1$ or 2 for a 1-ended or 2-ended loss estimate (whether we count losses at one or both ends of the field line), $L_{\parallel}\sim 1$m is a characteristic connection length, and $K_{i}(T_e)$ is the ionization rate coefficient, for which the Voronov fit\cite{voronov1997practical} gives $K_i=5.3\times 10^{-15}$m$^3$s$^{-1}$ at 10 eV and $K_i=1.4\times 10^{-14}$m$^3$s$^{-1}$ at 20 eV. The neutral density estimated from (\ref{estn0}) is shown in figure \ref{fig:ionfrac}. It does exceed $10^{19}$m$^{-3}$ for sufficiently low $L_{\parallel}$ and $T_e$. Low density operation may mean that the LH resonance is not inside the private SOL, but rather in front of the antenna where $L_{\parallel}\gg 1$m, in which case the neutral density should be much lower.}

\textcolor{rev0}{Experimental data on the topic of the neutral density is scarce. The neutral \emph{pressure} is sometimes measured in the edge\cite{labombard2000crossfield,klepper1993neutral,kallenbach2019neutral,scarabosio2009measurements,zhou2023measurement,diab2026direct,wang2007neutral}, but never in the antenna box, and without knowledge of the neutral temperature there remains substantial uncertainty on the corresponding neutral density. A rare example of a direct neutral density measurement is\cite{boivin2001neutral}, reporting $n_0=$2-3$\times 10^{17}$m$^{-3}$ at a location 5-10 mm outside the separatrix in Alcator C-Mod.
}

Maquet et al\cite{maquet2021analytical} give the width of the near-singular electric field at the Lower Hybrid resonance, which takes a Lorentzian form in the collisional case (finite but sharply peaked), as
\begin{align}
    W_{LH}=\frac{2\Im(S)}{\left|\Re\left(\frac{dS}{dx}\right)\right|} \label{WLH}
\end{align}
where the $x$-dependent Stix parameter\cite{stix1992waves} $S(x)$ and its derivative are evaluated at $\Re(S(x))=0$. If the edge density grows exponentially as $\exp(x/\lambda)$, then $\left|\Re\left(\frac{dS}{dx}\right)\right|=1/\lambda$ at $\Re(S(x))=0$. So (\ref{WLH}) is the product of a physics-set dimensionless factor $2\Im(S)$, and a plasma profile-set length scale $\lambda$.

\begin{figure}
    \centering
    \includegraphics[width=0.75\linewidth]{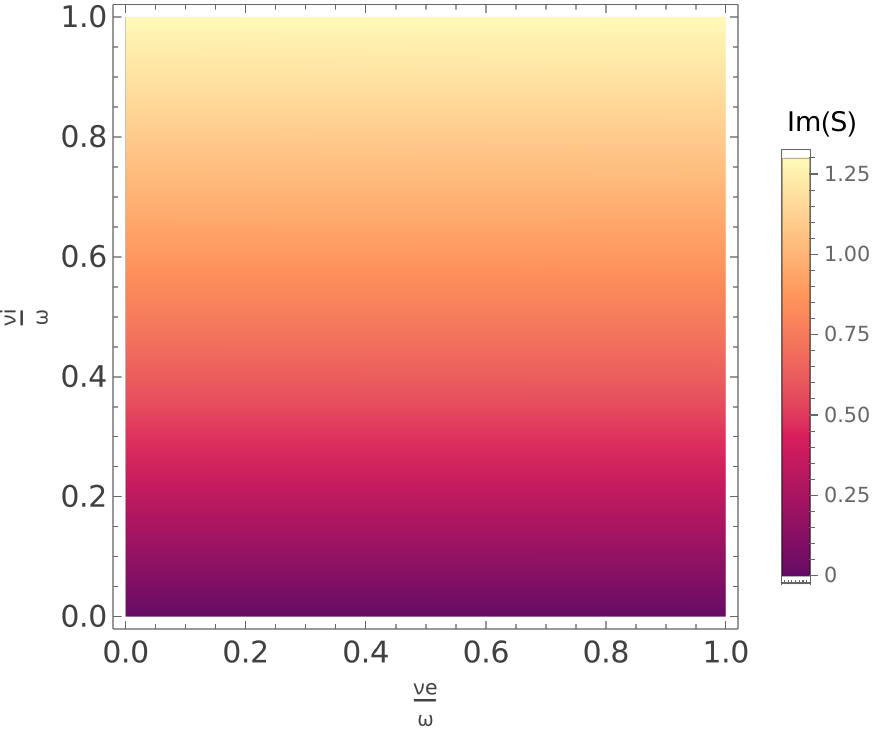}
    \caption{$\Im(S)$ at $\Re(S)=0$ for various ion and electron collision frequencies. The magnetic field is 2T, the wave frequency is $\omega=2\pi \cdot (30\text{ MHz})$.}
    \label{fig:imS}
\end{figure}

At $\Re(S)=0$,
\textcolor{rev0}{\begin{align}
    \Im(S) &= \frac{\nu_e}{\omega} \frac{m_e}{m_i} \left(\frac{\omega ^2-\Omega_i^2}{ \Omega_i^2} +O\left(\frac{m_e}{m_i}\right)\right)+O\left(\left(\frac{\nu_e}{\omega}\right)^2\right)\nonumber \\
   &+\frac{\nu_i}{\omega} \left(\frac{\omega ^2+\Omega_i^2}{\omega ^2-\Omega_i^2}+O\left(\frac{m_e}{m_i}\right)\right) +O\left(\left(\frac{\nu_i}{\omega}\right)^2\right) \label{eqImS}
\end{align}
(\ref{eqImS}) is typically dominated by its second term $\propto\nu_i$. The first term $\propto\nu_e$ is suppressed by $m_e/m_i$. As such, it is mainly ion-neutral collisions that drive the width of the collisional Lower Hybrid resonance, the ion-electron Coulomb collision contribution to $\nu_i$ being suppressed by another factor $m_e/m_i$.}
We show $\Im(S)$ in figure \ref{fig:imS}.

\begin{figure}
    \centering
    \includegraphics[width=0.75\linewidth]{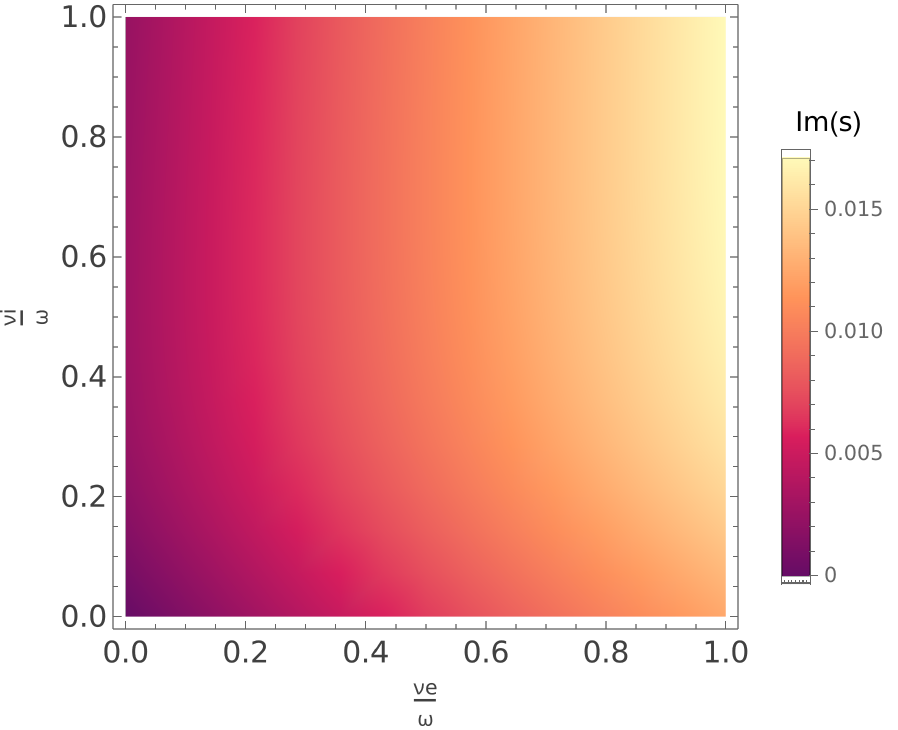}
    \caption{$\Im(s)=\Im(\sqrt{-S/P})$ at $n_e=10^{16}$m$^{-3}$, for various ion and electron collision frequencies. Because $|P|\gg|S|$ in most of the regime where the resonance cone exists, $\Im(s)$ tends to be dominated by $\Im(P)$ and hence by electron collisions, in sharp contrast with the ion collisionality dominated $\Im(S)$ from figure \ref{fig:imS}.}
    \label{fig:imsRC}
\end{figure}

\begin{figure}
    \centering
    \includegraphics[width=0.75\linewidth]{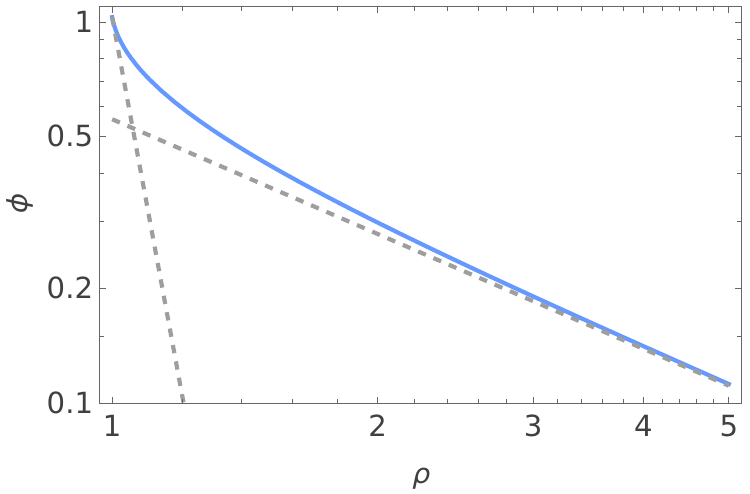}
    \caption{Transition between near-field and far-field behavior in the resonance cone. The potential (\ref{potentialAlongLine}) is shown in blue. Its asymptotes (\ref{asymInf}) and (\ref{asym1}) are shown as gray dashed lines. The intersection between these asymptotes is given by (\ref{asympIntersect}).}
    \label{fig:resonanceConeAsymp}
\end{figure}

In \cite{tierens2024slow}, we gave the analytic solution for resonance cone emission from a cylindrical source normal to $\boldsymbol{B}$. The nearly electrostatic slow wave obeys $\nabla\cdot\epsilon\nabla\phi=0$ or 
\begin{align}
    P \frac{\partial^2}{\partial_z^2} \phi(x,z)+S \frac{\partial^2}{\partial_x^2} \phi (x,z)=0 \label{slowWavePDE}
\end{align}
If the potential along the source boundary (radius $r_s$) is $A_m\exp(-i m \theta)$, the potential (solution of (\ref{slowWavePDE})) is
\begin{align}
    \phi=A_m F_{-m}\left(\frac{x+ s z}{r_s}\right) \label{RCcircSol}
\end{align}
with $s=\sqrt{-S/P}$ (complex with $\Re(s)>0$ and $\Im(s)>0$ in the collisional case) and
\begin{align}
    F_{-m}(\xi)=\left(\frac{\xi\left(\sqrt{1-\frac{s^2+1}{\xi^2}}-1\right)}{\sqrt{-s^2}-1}\right)^m
\end{align}
Like the lower hybrid resonance at $S=0$, the resonance cone electric field $-\nabla\phi$ has, in the collisionless limit, a singular component normal to the source surface, singular at those points where the cylinder is tangent to the resonance cone.
For $m=1$, the potential along a line normal to the surface at these tangency points is
\begin{align}
    \phi(\rho)=\frac{A_m\rho( s \Re(s)+1) \left(\sqrt{1-\frac{\left(s^2+1\right) \left(\Re(s)^2+1\right)}{\rho( s
   \Re(s)+1)^2}}-1\right)}{\sqrt{\Re(s)^2+1}} \label{potentialAlongLine}
\end{align}
where $\rho=r/r_s$. A series expansion around $\rho=\infty$ is
\begin{align}
    |\phi(\rho)|=\frac{1}{\rho}\left|\frac{A_m\left(s^2+1\right) \sqrt{\Re(s)^2+1}}{2 (s \Re(s)+1)}\right|+O\left(\left(\frac{1}{\rho}\right)^3\right) \label{asymInf}
\end{align}
The behavior near $\rho=1$ is
\begin{align}
    |\phi(\rho)|=|A_m(s-i)| \rho^{-\frac{\Re(s)^2+1}{\Im(s)}} \label{asym1}
\end{align}
The singular derivative (singular normal electric field) as $\Im(s)\rightarrow 0$ is especially clear from (\ref{asym1}).
The transition between near-field behavior and far-field behavior occurs at
\begin{align}
    \rho= \left| \frac{(s+i) \sqrt{\Re(s)^2+1}}{2(s \Re(s)+1)}\right|
   ^{\frac{1}{\frac{-\Re(s)^2-1}{\Im(s)}+1}}  \label{asympIntersect}
\end{align}
This is shown in figure \ref{fig:resonanceConeAsymp}. Together with the source radius of curvature, (\ref{asympIntersect}) defines a length scale which needs to be resolved in order to handle the resonance cone in cold plasma. By virtue of being defined as an intersection of straight lines in a log-plot, for any given mode $m$, this length scale depends neither on the amplitude $A_m$ (which would multiply the potential, hence a vertical translation on the log-scale, which leaves the intersection invariant) nor on the mode number $m$ (which would raise the potential to a power, hence a vertical scaling on the log-scale, which also leaves the intersection invariant), though technically linear combinations of modes do not leave this length scale invariant.
(\ref{asympIntersect}) is singular at $\Im(s)=\Re(s)^2+1$. The resonance cone occupies the density interval between the $P=0$ cutoff and the $S=0$ lower-hybrid resonance, on which $S$ is bounded and $\Im(s)$ remains below this singular curve, reaching it only at $s=i$ in the unphysical limit of infinite collision frequencies.
Unlike $\Im(S)$ from figure \ref{fig:imS}, which depended chiefly on the ion collisionality, $\Im(s)$ shown in figure \ref{fig:imsRC} depends mostly on the electron collisionality.

We do not, in this work, make use of sheath boundary conditions (using perfectly conducting walls instead), but we do note that the length scale defined by (\ref{asympIntersect}) is often shorter than the sheath width (some rectification-dependent multiple of the Debye length). It is generally accepted \cite{myra2023validity} that sheath boundary conditions need modifications at $\boldsymbol{B}$ tangency points, those points at which the confining magnetic field is tangent to the plasma-facing component surface, because in that case the electrons are not free to flow perpendicular to the plasma-facing component surface. (\ref{asympIntersect}) suggests another, previously undocumented, failure mode of the sheath boundary condition, a failure of scale-separation. At resonance cone tangency points ($\neq\boldsymbol{B}$ tangency points), the geometry- and collisionality-dependent perpendicular scale (\ref{asympIntersect}) may be substantially smaller than the sheath width. The effect of that is beyond the scope of this work.

\begin{figure}
    \centering
    \includegraphics[width=0.95\linewidth]{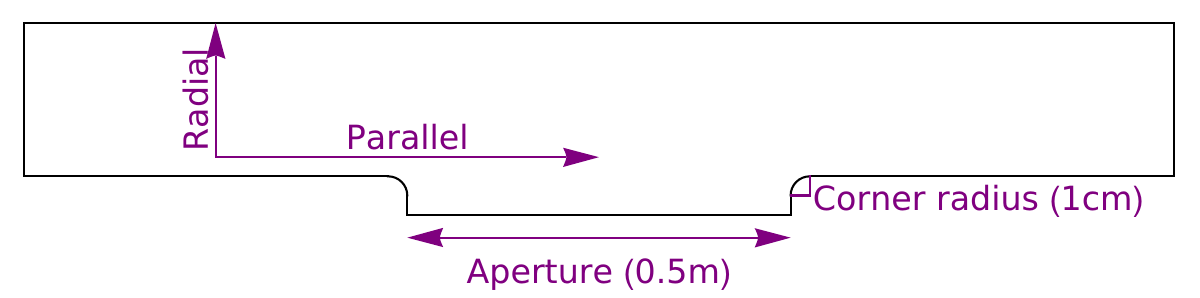}
    \caption{2D Finite Element simulation domain of an ICRF antenna in a low-density edge plasma.}
    \label{fig:antDiag}
\end{figure}

\begin{figure}
    \centering
    \includegraphics[width=0.49\linewidth]{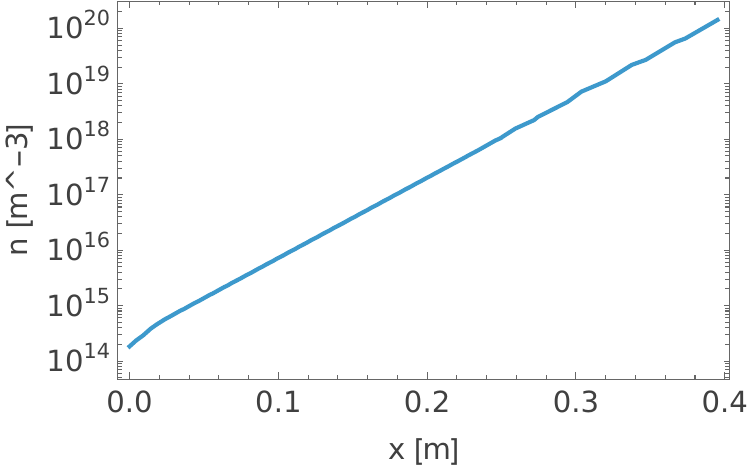}
    \includegraphics[width=0.49\linewidth]{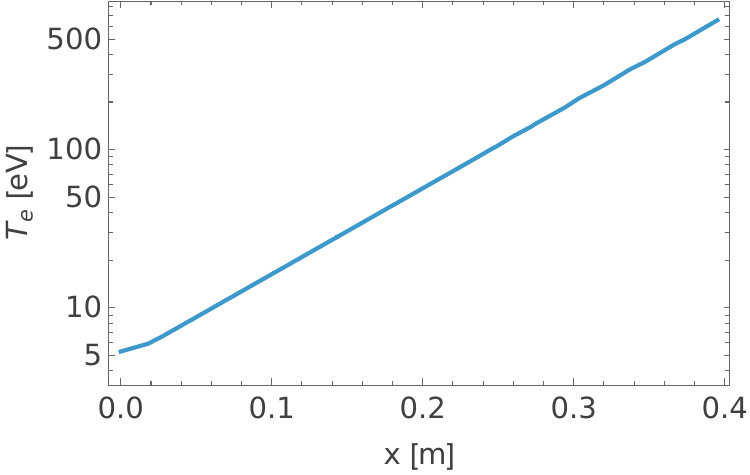}
    \caption{Density and temperature for our low-density edge plasma.}
    \label{fig:nete}
\end{figure}

To illustrate these results, we have constructed a 2D Finite Element simulation shown in figure \ref{fig:antDiag}. An anti-symmetric source current in the out-of-plane (poloidal) direction is imposed at the aperture. A perfectly matched layer absorbs the waves at the core side of the simulation domain. The magnetic field strength is $B=2$T, the wave frequency is $f=30$MHz, the density and temperature are in figure \ref{fig:nete}. This is an unusually low edge density for today's devices.

Under these assumptions, the relevant length scales (\ref{WLH}) and (\ref{asympIntersect}) are summarized in table \ref{tab:power}, for various neutral densities. At higher neutral densities, the collision frequencies rapidly become dominated by elastic electron-neutral collisions and by ion-neutral charge exchange \textcolor{rev0}{(assuming a D rather than D$_2$ neutral population)}, the reaction \textcolor{rev0}{rates being $\sim 10^{-13}$m$^3$/s and $\sim 10^{-14}$m$^3$/s respectively}. We encounter the regime in which both $2\Im(S)$ (the dimensionless factor in the LH scale) and $\rho-1$  (the dimensionless factor in the RC scale) are near-linear in the neutral density. These dimensionless factors remain very small at realistic neutral densities. For the density length scale $\lambda$ ($n_e\propto \exp(x/\lambda)$) of order 10cm, the LH width $2\Im(S)\lambda$ exceeds the electron Larmor radius ($\sim 10\mu$m at $B=2$T and $T_e=10$eV) for neutral densities $>10^{18}$m$^{-3}$. The situation is worse yet for the resonance cone: for a corner radius of curvature of order $r_s\sim 1$cm, $(\rho-1)r_s$ exceeds the electron Larmor radius only at neutral densities above $10^{20}$m$^{-3}$.

\begin{figure}
    \centering
    \includegraphics[width=0.32\linewidth]{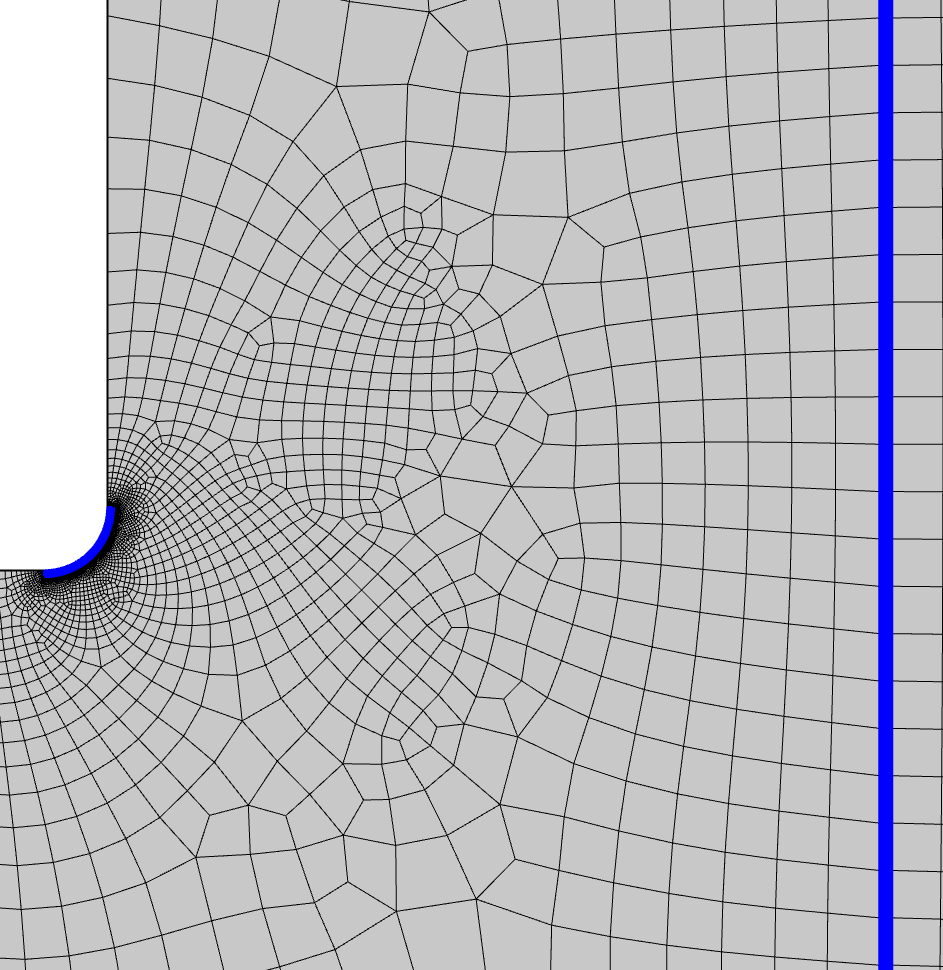}
    \includegraphics[width=0.32\linewidth]{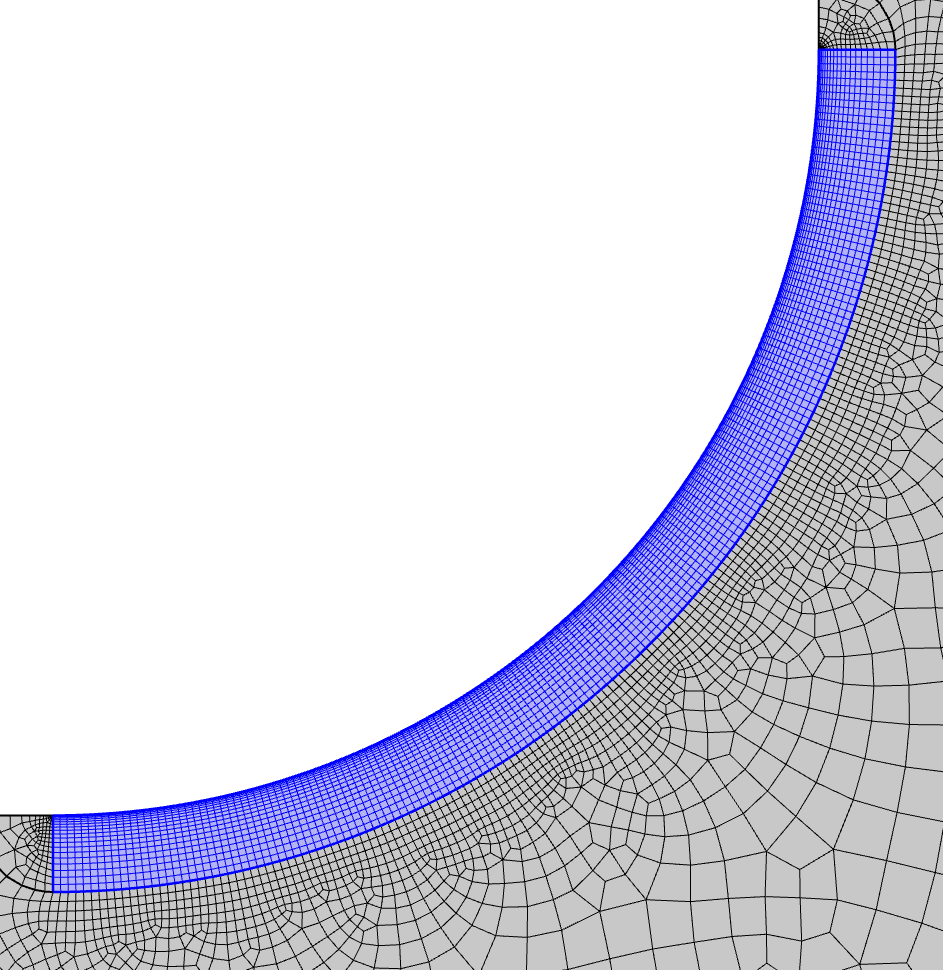}
    \includegraphics[width=0.32\linewidth]{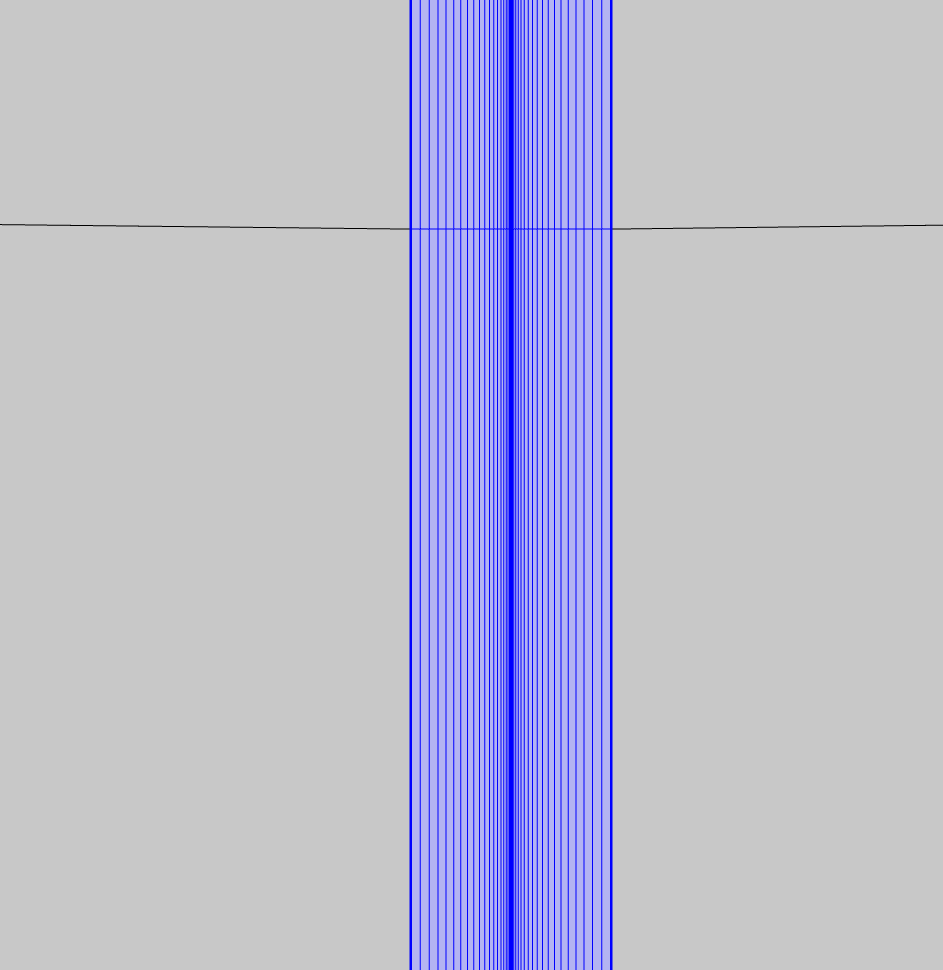}
    \caption{Exponential mesh spacing resolves both the LH resonance and the resonance cone.}
    \label{fig:expmesh}
\end{figure}

\begin{figure}
    \centering
    \includegraphics[width=1.0\linewidth]{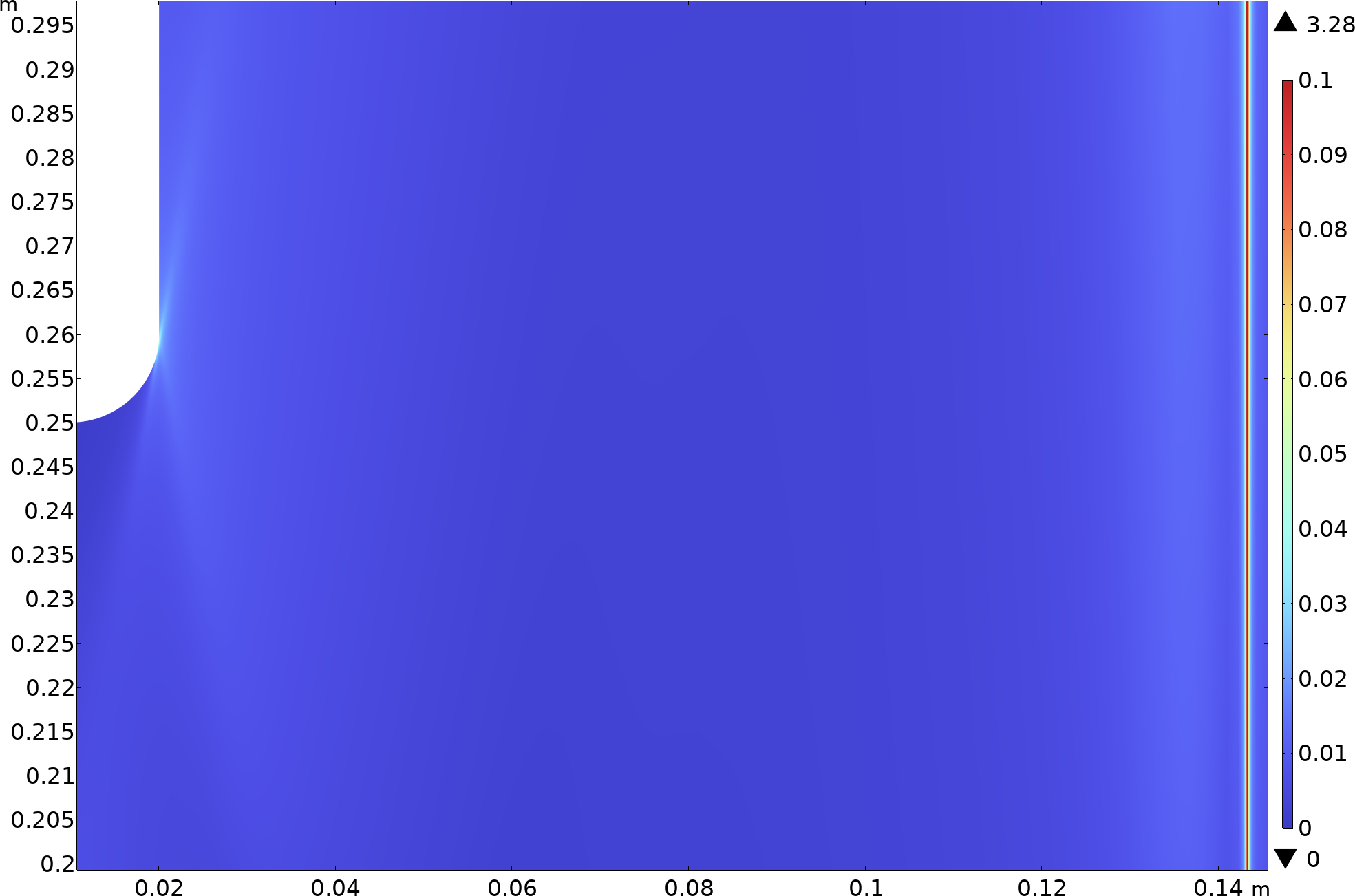}
    \caption{In-plane electric field norm in arbitrary units. Note that we see both predicted phenomena: the resonance cone excited through mode conversion at the tangency point (left), and the lower hybrid resonance at $S=0$ (right, at $x\approx 14$cm).}
    \label{fig:normE}
\end{figure}

\begin{figure}
    \centering
    \includegraphics[width=0.45\linewidth]{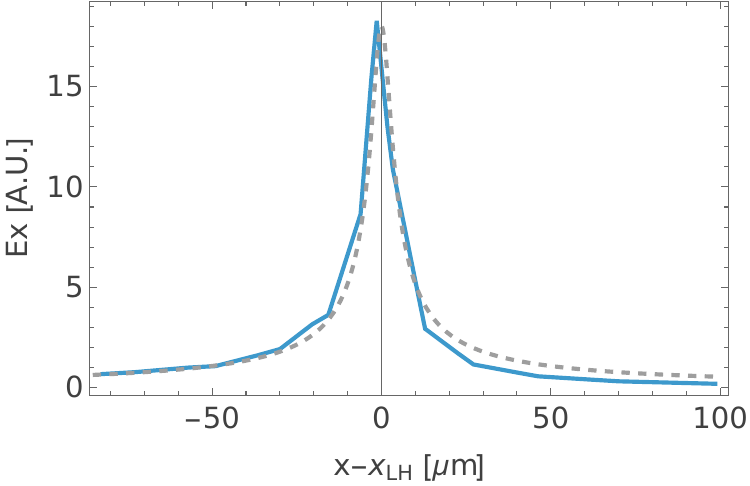}
    \includegraphics[width=0.45\linewidth]{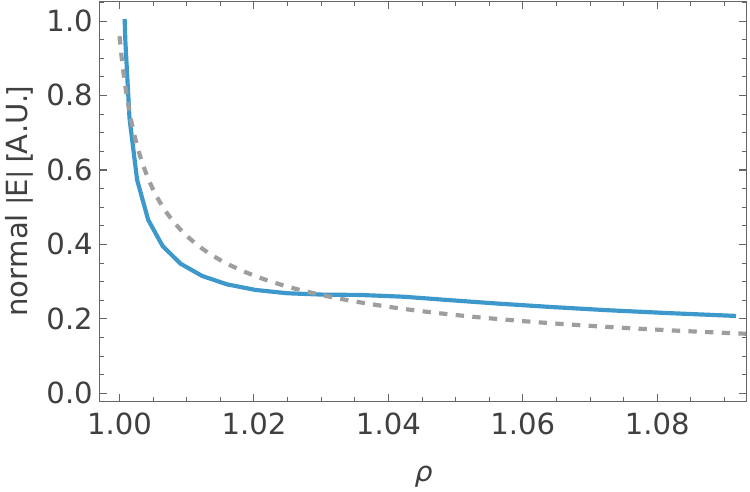}
    \caption{Left: $|E_x|$ near the lower hybrid resonance.  Right: $|E_n|$ (normal to the quarter-circle limiter corner) at the resonance cone tangency point. Blue: numerical Finite Element solution. Gray dashed: corresponding analytic result.}
    \label{fig:analyticCompare}
\end{figure}

Despite these short length scales, at sufficient neutral density ($n_n=10^{20}$m$^{-3}$), we do succeed in resolving them thanks to the use of exponential mesh spacing, as shown in figure \ref{fig:expmesh}. Figure \ref{fig:normE} confirms that both phenomena (resonance cone, lower hybrid resonance) show up in the numerical solution. Figure \ref{fig:analyticCompare} confirms that both behave as analytically predicted.

\begin{table}
    \centering
    \begin{tabular}{c|c|c|c}
         Collisions & Neutral density & $2\Im(S)$ & $\rho-1$ \\ \hline
         Coulomb+neutral & $10^{16}$m$^{-3}$ & $2\times 10^{-6}$ & $9\times 10^{-7}$ \\
         Coulomb+neutral & $10^{17}$m$^{-3}$ & $2\times 10^{-5}$ & $3\times 10^{-6}$ \\
         Coulomb+neutral & $10^{18}$m$^{-3}$ & $2\times 10^{-4}$ & $3\times 10^{-5}$ \\
         Coulomb+neutral & $10^{19}$m$^{-3}$ & $2\times 10^{-3}$ & $3\times 10^{-4}$ \\
         Coulomb+neutral & $10^{20}$m$^{-3}$ & $2\times 10^{-2}$ & $3\times 10^{-3}$ \\
         Coulomb+neutral & $10^{21}$m$^{-3}$ & $2\times 10^{-1}$ & $3\times 10^{-2}$ \\
    \end{tabular}
    \caption{Summary of the length scales (\ref{WLH}) and (\ref{asympIntersect}) associated with the near-resonant phenomena in our numerical example. The dimensionless parameter $2\Im(S)$ at $\Re(S)=0$, multiplied by the density length scale $\lambda$ ($n_e\propto \exp(x/\lambda)$), gives the width of the collisional Lower Hybrid resonance. The dimensionless parameter $\rho-1$, multiplied by the corner radius, gives the resonance cone length scale.}
    \label{tab:power}
\end{table}

In this work, we have investigated the length scales associated with the resonance cone and the lower hybrid resonance under physical collisions. While it is true that the mathematical singularities of the collisionless theory are removed, and replaced by sharply peaked but finite fields, the length scales associated with those not quite singular field peaks remain governed by small dimensionless factors. For the lower hybrid resonance, this factor is $2\Im(S)$, which is chiefly determined by the \textcolor{rev0}{ion-neutral} collision frequency. For the resonance cone, it is $\rho-1$, which is chiefly determined by the electron collision frequency.
\textcolor{rev0}{The neutral density is thus not
a peripheral input: it directly sets the regularizing length scales that determine whether a collisional cold plasma
Finite Element calculation is numerically meaningful. This neutral density in the antenna private scrape-off layer is presently not well constrained. In particular, extremely low ionization $n_e/n_0\lesssim 10^{-3}$ (necessary to achieve MHz collision frequencies) at the LH density should not be excluded solely from thermodynamic arguments, because the plasma in this region is not in thermodynamic equilibrium.}

The \emph{physical} adequacy of the collisional cold plasma description remains a separate question. In this low-density regime, the rectified sheaths become macroscopic, a point routinely neglected when modeling the sheath as a boundary condition. These wide sheaths may interact non-conservatively with electrons (``Fermi acceleration''\cite{lieberman1998fermi} or stochastic heating) and may cause the electron temperature to rise well above its thermal equilibrium value, making collisional cold plasma (and possibly even Maxwellian hot plasma) an inadequate description. Whether this occurs at ICRF frequencies inside antenna structures is not established. We plan to investigate this on WEST in future work.


\nocite{*}
\bibliography{aipsamp}

\end{document}